\documentclass[]{aastex631}

\usepackage{epstopdf}

\begin{document}

\title{Category-based Galaxy Image Generation via Diffusion Models}

\correspondingauthor{Hongming Tang}
\email{Hongming.Tang@xjtlu.edu.cn}

\author[0009-0000-4781-5395]{Xingzhong Fan}
\affiliation{Department of Physics, Xi'an Jiaotong-Liverpool University, Suzhou 215123, China}

\author[0000-0002-7300-9239]{Hongming Tang}
\affiliation{Department of Physics, Xi'an Jiaotong-Liverpool University, Suzhou 215123, China}

\author[0009-0001-4759-9968]{Yue Zeng}
\affiliation{Department of Computer Science, University of Illinois at Urbana-Champaign, Champaign, IL 61820, USA}

\author[0000-0002-1805-0570]{M.B.N. Kouwenhoven}
\affiliation{Department of Physics, Xi'an Jiaotong-Liverpool University, Suzhou 215123, China}

\author[0000-0003-4708-3344]{Guangquan Zeng}
\affiliation{Department of Physics, The Chinese University of Hong Kong, Sha Tin, N.T., Hong Kong, China}






\begin{abstract}

Conventional galaxy image generation methods rely on semi-analytical models and hydrodynamic simulations, which are highly dependent on physical assumptions and parameter tuning. In contrast, data-driven generative models do not have explicit physical parameters pre-determined, and instead learn them efficiently from observational data, making them alternative solutions to galaxy generation. Among these, diffusion models outperform Variational Autoencoders (VAEs) and Generative Adversarial Networks (GANs) in quality and diversity. Embedding generalized physical features, such as category information, further enhances their generative capabilities. In this work, we present GalCatDiff, the first framework in astronomy to leverage both galaxy image features and astrophysical properties in the network design of diffusion models. GalCatDiff incorporates an enhanced U-Net and a novel block entitled Astro-RAB (Residual Attention Block), which dynamically combines attention mechanisms with convolution operations to ensure global consistency and local feature fidelity. Moreover, GalCatDiff uses category embeddings for class-specific galaxy generation, avoiding the high computational costs of training separate models for each category. Our experimental results demonstrate that GalCatDiff significantly outperforms existing methods in terms of the consistency of sample color and size distributions, and the generated galaxies are both visually realistic and physically consistent. This framework will enhance the reliability of galaxy simulations and can potentially serve as a data augmentor to support future galaxy classification algorithm development.
\end{abstract}




\section{Introduction} 
\label{sec:intro}
Galaxies exhibit a wide range of morphologies \citep[e.g.,][]{hubble1926galaxy, sandage1961hubble, vandenbergh1976galaxy, kormendy2012galaxy, willett2013galaxyzoo, buta2015galaxy, walmsley2022galaxy, walmsley2023gzdesi}.
The morphology of galaxies is closely linked to various physical properties, reflecting the cumulative effects of different physical processes throughout their evolutionary history \citep[e.g.,][]{buta2013galaxymorphology, conselice2014galaxy, cappellari2016ifu}. 
For example, spiral galaxies tend to be bluer, exhibit prominent star formation, and are rotation-dominated; in contrast, elliptical galaxies are generally redder and dominated by velocity dispersion \citep[e.g.,][]{strateva2001color, baldry2004colormangitude, kelvin2014morphology, zhu2024manga}.
In addition, it has been shown that disk galaxies with different bulge components, such as classical bulges or pseudobulges, exhibit significant differences in their disk properties, implying a co-evolution between the bulge and the whole galaxy \citep[e.g.,][]{ 2024A&A...691A.125H,2024MNRAS.529.4565H}.

Treating galaxy morphology as a proxy for galaxy diversity enables further investigation of the temporal evolution using redshifts extending to the early universe \citep[e.g.,][]{conselice2014evolution, 2024A&A...685A..48H, geron2025galaxy}. A recent study, for instance, analyzed model predicted galaxy morphology (Spheroids, Disks, Irregular and Bulge+disk) and redshift of $\sim$ 20,000 galaxies in the JWST \citep[James Webb Space Telescope,][]{gardner2023jwst} Cosmic Evolution Early Release Science survey \citep[CEERS,][]{finkelstein2022long,finkelstein2023ceers}, confirmed that bulge-dominated galaxies decrease with redshifts, while peculiar galaxies increased with redshifts \citep[][]{2024A&A...685A..48H}. In general, detailed morphological studies, especially accurate galaxy morphology classification over a large and comprehensive sample, should provide powerful diagnostic for understanding galaxy formation and evolution.

Classifying galaxy morphologies hence becomes a fundamental and crucial task for these studies.
The advent of large-scale imaging surveys now could provide millions of galaxy images \citep[e.g.,][]{2000AJ....120.1579Y, 2007ApJS..172...38S, 2009ApJS..182..543A, 2016MNRAS.460.1270D}.
Detailed morphological classifications of these galaxies not only enable the construction of useful statistical samples \citep[e.g.,][]{2018MNRAS.476.3661D, 2021MNRAS.507.4425C, 2021MNRAS.506.1927V}, but also facilitate the discovery of rare galaxy populations, such as superthin galaxies \citep[e.g.,][]{2014ApJ...787...24B, 2017MNRAS.465.3784B}.
However, the vast scale of these images presents a significant classification challenge.
While some efforts continue through manual inspection by professional astronomers \citep[e.g.,][]{2010ApJS..186..427N, 2011A&A...532A..74B, 2015ApJS..217...32B} or citizen science volunteers \citep[e.g., Galaxy Zoo,][]{2013MNRAS.435.2835W, 2017MNRAS.464.4176W}, the majority of work now relies on automated methods.
Moreover, since billions of galaxies are expected to be discovered from ongoing and upcoming surveys, such as LSST \citep[][]{2019ApJ...873..111I}, JWST \citep[][]{gardner2023jwst}, Euclid \citep[][]{2025A&A...697A...1E}, and CSST \citep[][]{2025arXiv250704618C}, automated classification becomes an absolute necessity, while rendering manual classification nearly impossible.
Early automated methods such as Decision Trees \citep[e.g.,][]{owens1996using} and Locally Weighted Regression \citep[e.g.,][]{de2004machine} were able to handle small datasets \citep[e.g.,][]{bazell2001ensembles}, while deep learning algorithms are more suitable to handle massive datasets as they have complex architectures and are able to extract features from high-dimensional data \citep[e.g.,][]{dieleman2015rotation, 2023walmsleydesi,luo2025galaxy}. Galaxy Zoo DESI, for instance, managed to perform detailed morphology measurements for 8.7 million galaxies in the DESI Legacy Imaging Surveys using a variant of EfficientNetB0 model trained on Galaxy Zoo DECaLS labeled data \citep{2023walmsleydesi}. 



Although deep learning has achieved great progress in galaxy morphology research, model training relies heavily on the availability of unbiased, high-quality labeled training data. Well-labeled data usually rely on manual identification, performed by either experts or citizen scientists. Both approaches require top-down designation from galaxy sample selection, tagging workflow, and validation procedure, which inevitably introduce human biases \citep[e.g.,][]{lintott2008galaxy, bamford2009galaxy, willett2013galaxyzoo}. Moreover, even if these biases can be partially resolved through careful selection of model architecture and training strategies \citep[e.g.,][]{walmsley2020galaxy,jiang2023galaxy}, data complexity of the less representative sample class may still be incomparable with those that dominate the total sample.

Deep generative models, however, can directly learn the underlying statistical features from large observational datasets without relying on explicit physical parameter settings  \citep[e.g.,][]{kingma2013VAE,goodfellow2014generative,ho2020denoising}. Such data-driven approaches accurately capture data features and leverage the capabilities of neural networks and parallel computing to achieve rapid generation \citep[e.g.,][]{ramos2019stokes,holdship2021chemulator,lanusse2021deep,smith2022realistic}, effectively addressing the limitations of conventional methods.

In recent years, deep generative models based on Variational Autoencoders (VAEs) and Generative Adversarial Networks (GANs) have already been applied to galaxy image generation tasks \citep{ravanbakhsh2017enabling,fussell2019forging,smith2019generative,lanusse2021deep,holzschuh2022realistic}. 
As a representative VAE-based algorithm, \cite{ravanbakhsh2017enabling} introduced a conditional VAE \citep[C-VAE;][]{kingma2014semi,sohn2015learning} for galaxy image generation, where the model is conditioned on the galaxy’s half-light radius, magnitude, and redshift. Their results demonstrated that, when measured from the generated images, the samples reproduce morphological statistics such as ellipticity without significant bias. However, the images tend to lack fine structural details, appearing relatively blurred compared to real observations. Compared with VAEs, GAN-based models are found to be able to generate more realistic galaxy mock images \citep{holzschuh2022realistic}. The limitation of GAN-based models, on the other hand, is that the generated images are inherently biased toward the training set, and large-size outputs become unrepresentative of true cosmological structures because rare features absent from the training data cannot be synthesized \citep[e.g.,][]{smith2019generative}. Additionally, training GANs remains challenging, with stable results achieved only when limited to three photometric bands \citep[e.g.,][]{smith2019generative}. In summary, VAEs struggle to produce high-quality images \citep[e.g.,][]{van2017neural,vahdat2020nvae}, while GANs often suffer from training instability, mode collapse, and lack of diversity in the generated samples \citep[e.g.,][]{goodfellow2014generative,salimans2016improved,2016arXiv161202136C}. Nevertheless, many GAN variants have sought to address these shortcomings; for example, StyleGAN has been applied to galaxy image generation and alleviates mode collapse to some extent \citep[e.g.,][]{holzschuh2022realistic}.

Diffusion models \citep[e.g.,][]{ho2020denoising} perhaps could serve as a better backbone when developing deep learning based galaxy image generative models. This is a recently developed generative model backbone that received popularity in a variety of research fields such as protein structure prediction and precipitation nowcasting. Researchers in these fields have explored leveraging physical priors into diffusion models to achieve physically consistent generation processes, further enhancing the model's capabilities \citep[e.g.,][]{abramson2024accurate,gao2024prediff}. In the broader machine learning community, advanced efforts such as Stable Diffusion \citep{rombach2022high} and Denoising Diffusion GANs \citep{xiao2021tackling} have been developed, which extend diffusion models by operating in latent spaces for efficiency or by integrating adversarial training to accelerate sampling. These advances demonstrate the versatility of diffusion models. However, astronomy-specific applications of these diffusion backbones remain at an early stage \citep[e.g.,][]{smith2022realistic,zhao2023can,reddy2024difflense}. \citet{zhao2023can}, for instance, demonstrates that diffusion models can support the conditional generation of astrophysical images and outperform StyleGAN \citep{karras2020analyzing}, one of the most robust GAN models in their experiments, producing more robust outputs while offering reliable parameter constraints. In realistic mock galaxy image generation, \citet{smith2022realistic} first trained a diffusion model and managed to generate images with visually realistic morphologies and consistent physical properties such as galaxy sizes and flux. 

Although diffusion models seem to shed light on the field of galaxy generation, \citet{smith2022realistic} had also pointed out that the computation cost of applying such a backbone is considerably higher than VAEs and GANs. Modifications to lower computation costs are then necessary. Moreover, we believe a sub-optimal diffusion-based model for galaxy generation should also be able to extract features at different scales so that galaxies with various distances, apparent angular size, morphology, and correlation with their surrounding environment can be generated. Developing and introducing a novel machine learning block to overcome these issues can largely resolve these issues, which is part of the study presented in this paper.

If such an algorithm can be developed, we further ask whether it can generate realistic mock galaxy images in a given morphological class. This is largely due to the data challenge that we developers are facing. In recent years, machine learning has been widely applied to handle galaxy-related downstream tasks such as source detection \citep[e.g.,][]{abraham2018detection,angora2020search,ciprijanovic2021deepmerge,sanchez2023identification}, classification \citep[e.g.,][]{dieleman2015rotation,aniyan2017classifying,cheng2020optimizing,walmsley2022galaxy,xu2023images} and segmentation \citep[e.g.,][]{boucaud2020photometry,ferreira2022simulation,arcelin2021deblending}. Although observational data are abundant, the distribution of different galaxy types in real datasets remains distinctly imbalanced. This imbalance directly affects the ability of machine learning models to learn minority classes, reducing their performance in these categories \citep[e.g.,][]{dieleman2015rotation,abraham2018detection,xu2023images}. Labelling galaxy types itself can also be time-consuming and possibly survey-dependent: a large, annotated galaxy morphology model training sample can take several years even with the support from citizen scientists \citep[e.g., Galaxy Zoo DECaLS;][]{walmsley2022galaxy}. A powerful, class-embedded deep generative model can serve as an efficient data augmenter, producing realistic mock images in customized class and enhancing the robustness of downstream machine learning tasks  \citep[e.g.,][]{holzschuh2022realistic,zhao2023can}.

Given this background, we carry out this proof-of-concept study by proposing a novel data-driven framework, GalCatDiff, a diffusion-based model that supports image generation conditioned on galaxy categories. By introducing category embeddings, GalCatDiff can generate images for specific galaxy types, rather than relying on indiscriminate sampling from the overall data distribution, thus avoiding the high cost of training separate generative models for each category. In standard diffusion models, the U-Net primarily consists of convolutional blocks for local feature extraction, with self-attention layers inserted at certain resolutions to complement convolutional blocks by modeling global interactions across feature maps \citep[e.g.,][]{ho2020denoising}. However, this design treats convolution and attention as separate blocks, without explicitly integrating their complementary strengths for joint global–local feature learning. Thus, we design an enhanced U-Net, which optimizes the backbone U-Net of the standard diffusion model, and introduce a module named Astro-Residual Attention Block (Astro-RAB). This module combines attention mechanisms with convolutional operations and dynamically adjusts their weights during network training, enabling the model to maintain overall consistency while accurately reproducing local features. Experimental results show that GalCatDiff outperforms existing methods in generating high-quality and physically consistent galaxy images, achieving significant progress.

The structure of this paper is as follows. Section~\ref{sec:ddpm} introduces the Denoising Diffusion Probabilistic Model \citep[DDPM;][]{ho2020denoising} we use in this work as the model backbone and the network architecture of GalCatDiff. Section~\ref{sec:experim} describes the dataset construction and experimental setup. Section~\ref{sec:result} presents the generation results and discussions. Finally, Section~\ref{sec:conclusion} presents the conclusions and outlook.


\section{Denoising Diffusion Probabilistic Models}
\label{sec:ddpm}

DDPM models \citep{ho2020denoising} provide a groundbreaking approach in generative modeling, leveraging Markov chains, and variational inference. In DDPMs, a forward diffusion process gradually corrupts the image with Gaussian noise until the original signal is annihilated, while the reverse process iteratively denoises corrupted inputs to generate high-quality samples. A schematic overview of the DDPM process is shown in the upper part of Fig.~\ref{fig:unet}.

The forward process assumes a Markov process, defined as \( q(\mathbf{x}_t \mid \mathbf{x}_{t-1}) \), where the image \( \mathbf{x}_0 \) is gradually transformed into isotropic Gaussian noise \( \mathbf{x}_T \) over \( T \) timesteps, which can be expressed as:
\begin{equation}
    q(\mathbf{x}_t \mid \mathbf{x}_{t-1}) 
    = \mathcal{N}(\mathbf{x}_t; \sqrt{1-\beta_t} \,\mathbf{x}_{t-1}, \beta_t \mathbf{I})
    \quad ,
\end{equation}
where $\beta_t$ is the noise variance schedule, and $\mathbf{I}$ is the identity matrix. By recursively applying the diffusion process, sampling $x_t$ at an arbitrary timestep $t$ in closed form:
\begin{equation}
    q(\mathbf{x}_t \mid \mathbf{x}_0) = 
    \mathcal{N}(\mathbf{x}_t; \sqrt{\bar{\alpha}_t} \mathbf{x}_0, (1 - \bar{\alpha}_t) \mathbf{I})
    \quad .
\end{equation}
Here, $\alpha_t = 1 - \beta_t$ denotes represents noise schedule at each timestep, and $\bar{\alpha}_t = \prod_{s=1}^t \alpha_s$ is the cumulative product of the noise schedule up to timestep $t$. Thus, \( \mathbf{x}_t(\mathbf{x}_0, \epsilon) = \sqrt{\bar{\alpha}_t} \mathbf{x}_0 + \sqrt{1 - \bar{\alpha}_t} \epsilon \), where \( \epsilon \sim \mathcal{N}(0, \mathbf{I}) \). This formulation ensures that as $t$ approaches final timestep $T$, the distribution of \( \mathbf{x}_T \) converges to a standard Gaussian distribution, \( \mathcal{N}(0, \mathbf{I}) \).

To generate the image, DDPMs learn the reverse process \( p_\theta(\mathbf{x}_{t-1} \mid \mathbf{x}_t, c) \), modeled as a parameterized Gaussian distribution. This reverse process is defined as:
\begin{equation}
    p_\theta(\mathbf{x}_{t-1} \mid \mathbf{x}_t, c) 
    = \mathcal{N}(\mathbf{x}_{t-1}; \mu_\theta(\mathbf{x}_t, t, c), \Sigma_\theta(\mathbf{x}_t, t, c))
    \quad ,
\end{equation}
where \( \mu_\theta(\mathbf{x}_t, t, c) \) represents the predicted mean of the Gaussian distribution at timestep $t$ with conditional variable $c$ (e.g., class label), used to iteratively denoise corrupted input and recover clean image. The expression for \( \mu_\theta(\mathbf{x}_t, t, c) \) is given by:
\begin{equation}
    \mu_\theta(\mathbf{x}_t, t, c) = \frac{1}{\sqrt{\alpha_t}} \left( \mathbf{x}_t - \frac{\beta_t}{\sqrt{1 - \bar{\alpha}_t}} \mathbf{\epsilon}_\theta(\mathbf{x}_t, t, c) \right)
    \quad ,
\end{equation}
where \( \mathbf{\epsilon}_\theta(\mathbf{x}_t, t, c) \) represents the model's predicted noise at timestep $t$, and $c$ is a conditional variable that guides the generation process. The covariance matrix \( \Sigma_\theta(\mathbf{x}_t, t) = \sigma_t^2 \mathbf{I} \) is experimentally assumed to be either $\sigma_t^2 = \beta_t$ or $\sigma_t^2 = \tilde{\beta}_t = (1-\bar{\alpha}_{t-1})(1-\bar{\alpha}_t)^{-1}$.

The reverse process starts at the final timestep, where \( \mathbf{x}_T \sim \mathcal{N}(0, \mathbf{I}) \), indicating that the image is fully corrupted by noise. From this point, the model progressively denoises the image until it reconstructs the original clean image \( \mathbf{x}_0 \). By conditioning this reverse process on the variable $c$, the model can generate images corresponding to specific categories, facilitating conditional image generation.

The objective of DDPMs is to maximize the data likelihood \( p_\theta(\mathbf{x}_0) \). Using variational inference, the training objective simplifies to minimizing the error in predicting the added noise, typically expressed as:
\begin{equation}
    L_\theta = \mathbb{E}_{\mathbf{x}_0, \epsilon, t, c} \left[ \left\| \epsilon - \mathbf{\epsilon}_\theta \left( \mathbf{x}_t, t, c \right) \right\|^2 \right]
    \quad ,
\label{eq:loss}
\end{equation}
with \( \mathbf{x}_0 \) as original clean image, \( \epsilon \sim \mathcal{N}(0, \mathbf{I}) \) as noise added in the forward process, $t$ is the time step, $c$ representing conditional input, and \( \epsilon_\theta(\mathbf{x}_t, t, c) \) is the predicted noise output by the U-Net model.


\subsection{Model Architecture}\label{subsec:arch}

In the inverse process of diffusion models, U-Net model is commonly used for predictions \citep[e.g.,][]{ho2020denoising, dhariwal2021diffusion, croitoru2023diffusion}. U-Net is a convolutional neural network (CNN)-based architecture initially designed for medical image segmentation tasks, focusing on achieving high-precision, pixel-level annotations in images \citep[e.g.,][]{ronneberger2015u,siddique2021u}.

A key feature of U-Net is its symmetric encoder-decoder structure \citep{NIPS2014_a14ac55a}. The encoder progressively downsamples the input image, extracting hierarchical feature representations, while the decoder incrementally restores the spatial resolution, ultimately producing the segmentation output. U-Net also incorporates skip connections, which directly pass feature maps from the encoder to corresponding layers in the decoder. These skip connections help the decoder retain high-resolution information, preventing the loss of detail during downsampling and upsampling, and allowing the model to better capture and reconstruct fine structural details \citep[e.g.,][]{mao2016image, drozdzal2016importance}.

\begin{figure}[htbp]
    \plotone{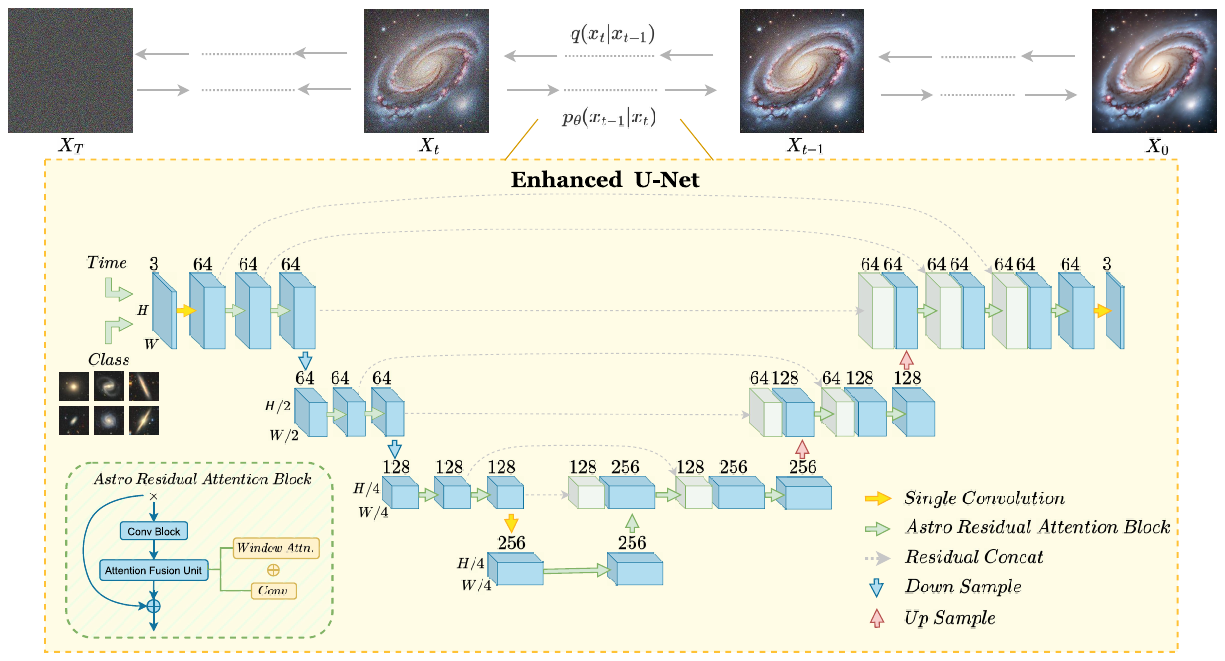}
    \caption{The schematic diagram of the diffusion model framework for category galaxy generation (GalCatDiff). In the forward process, a clean image $X_0$ is progressively corrupted by adding Gaussian noise at each timestep $t$, eventually producing a pure noise image $X_T$. The reverse process, guided by the enhanced U-Net estimating $P_\theta(X_{t-1} \mid X_t)$, starts from the noise image $X_T$ and iteratively removes noise to reconstruct the target image $X_0$. The enhanced U-Net architecture begins with a Single Convolution layer that transforms the input into a richer feature representation. The model consists of two down-sampling and up-sampling stages, each incorporating multiple Astro-RABs. During down-sampling, feature maps pass through each Residual Block, with skip connections established to retain key features and link them to corresponding layers in the up-sampling stages. Category and time information is integrated into the U-Net by embedding them and inputting them into each Astro-RAB. A schematic of the Astro-RABs is shown in the lower-left corner of this figure, and the full version is provided in Fig.~\ref{fig:resblock}.
    \label{fig:unet}}
\end{figure}

\begin{figure}[htbp]
    \plotone{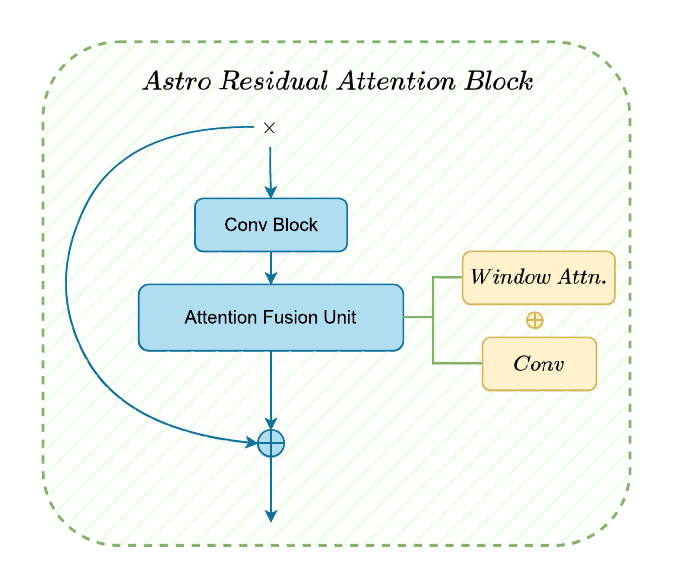}
    \caption{Overview of the Astro-RAB structure. Each block consists of a convolutional block followed by an Attention Fusion Unit and a skip connection, with this sequence iterated twice. The Attention Fusion Unit incorporates window attention, combined with convolutional layers, to preserve essential galaxy physics properties.
    \label{fig:resblock}}
\end{figure}

In the GalCatDiff framework, we propose an enhanced U-Net designed to generate high-quality galaxy images based on class labels, while preserving their physical properties, as shown in Fig.~\ref{fig:unet}. The main improvements include the introduction of the Astro-RAB, the use of multiple Astro-RABs (two or more) in each down-sampling and up-sampling stage, and additional skip connections from each convolutional or Astro-RAB layer in the down-sampling path to the corresponding layer in the up-sampling path, making the architecture wider than the standard U-Net. The additional skip connections facilitate the fusion of multi-scale features, reducing overfitting and thereby improving the model’s performance. The Astro-RAB integrates a fusion window attention \citep{liu2021swin} and convolution within the conventional residual block \citep{he2016deep} to preserve the physical characteristics of galaxies, as illustrated in Fig.~\ref{fig:resblock}.

The embedding process for galaxy categories consists of two main steps. First, an embedding layer is used to map the discrete category indices of galaxies into a fixed-dimensional continuous vector space, obtaining the initial category embedding representations. Subsequently, these embedding vectors are passed through a two-layer multilayer perception (MLP) for nonlinear transformation. The MLP includes two fully connected layers and a nonlinear activation function to enhance the feature representation capabilities. Through this process, the category indices are transformed into high-dimensional feature vectors containing galaxy category information, which are then input into each Astro-RAB for category embedding. Furthermore, the time step $t$ is embedded using a sinusoidal embedding (\cite{vaswani2017attention}), and input into each Astro-RAB for time embedding.

The overall architecture of our U-Net consists of two downsampling stages, a bottleneck stage, and two upsampling stages. Each downsampling and upsampling stage includes two Astro-RABs and a self-attention layer \citep{bahdanau2014neural, vaswani2017attention}. After passing through the Astro-RAB in the downsampling stages, feature maps are transmitted to the corresponding upsampling stages via skip connections. Therefore, at each level, there are at least two skip connections. The bottleneck stage consists of a single convolutional layer that expands the channels without reducing the feature map size as downsampling, followed by a Astro-RAB, a self-attention layer, and another Astro-RAB for the corresponding upsampling. Additionally, when an image enters the U-Net, it first passes through a single convolutional layer to increase the number of channels, allowing for richer representations. This feature map is then connected to the final upsampling stage via a skip connection. 

The Astro-RAB consists of a convolutional block followed by an Attention Fusion Unit and a skip connection, with this sequence repeated twice. Astro-RAB concatenates the class embedding and time embedding, maps them to the module's output dimensions through an MLP, and adds the result to the first sequence, thereby integrating the time and class embeddings into the feature map. The Attention Fusion Unit is composed of a convolution block and a window attention block. The convolutional layer efficiently extracts spatial features through local receptive fields and weight sharing, making it particularly effective for tasks involving local structures or patterns \citep[e.g.,][]{krizhevsky2012imagenet, simonyan2014very, szegedy2015going}. The attention mechanism, in contrast, dynamically adjusts the focus of information across the entire feature map, capturing long-range dependencies \citep[e.g.,][]{bahdanau2014neural,vaswani2017attention,oktay2018attention}. By combining the strengths of convolution and attention, the model can leverage both the local feature extraction capabilities of convolutions and the global information modeling capabilities of attention mechanisms \citep[e.g.,][]{liu2021swin,dai2021coatnet,pan2022integration,wu2021cvt,wang2021pyramid}, enhancing its ability to capture the physical features of galaxies. The output of the Attention Fusion Unit is the weighted sum of the convolutional and attention outputs. Both the convolutional weight $w_{\text{conv}}$ and the attention weight $w_{\text{attn}}$ are learnable parameters. Our U-Net model learns the optimal balance of these weights across different blocks to achieve the best performance, as discussed in further detail in Section~\ref{subsec:impactPPRB}.

\section{Dataset and Experimental Setup} 
\label{sec:experim}

\subsection{Data Collection}
\label{subsec:datacollection}

The dataset used in this study was derived from the Galaxy10 DECaLS dataset\footnote{https://astronn.readthedocs.io/en/latest/galaxy10.html}, which contains 17,736 samples with image size $256 \times 256$ pixels colored galaxy images (g, r and z bands) separated in 10 classes. The original calibrated fluxes in each band are provided in linear units of nanomaggies in the Legacy Surveys\footnote{https://www.legacysurvey.org/dr10/description/}\citep{dey2019overview}. In Galaxy10, the released images are generated by normalizing the nanomaggy fluxes in each band and mapping the resulting g, r, and z values to the 0–255 integer range used for the RGB renderings. We refined into the Galaxy6 DECaLS dataset, consisting of six categories: Round Smooth Galaxies, In-between Round Smooth Galaxies, Barred Spiral Galaxies, Unbarred Spiral Galaxies, Edge-on Galaxies without Bulges, and Edge-on Galaxies with Bulges. It comprises 14,497 images, distributed across the categories as 2,633, 2,308, 2,035, 4,412, 1,361, and 1,748 images. The dataset was split into training and testing subsets in an 80:20 ratio \citep{joseph2022optimal}, providing ample data for the model to learn robust features during training while preserving a representative portion for evaluating its performance on unseen data. The detailed counts of images per category in the training, testing, and total sets are reported in Table~\ref{tab:dataset_stat}.

\begin{table}[htbp]
\centering
\caption{Counts by Category in Train, Test, and Total Sets.}
\resizebox{\textwidth}{!}{
\begin{tabular}{lccccccc}
\hline
\textbf{Dataset} & \textbf{Round Smooth} & \textbf{In-between Round Smooth} & \textbf{Barred Spiral} & \textbf{Unbarred Spiral} & \textbf{Edge-on without Bulge} & \textbf{Edge-on with Bulge} & \textbf{Total} \\
\hline
Train & 2106 & 1846 & 1628 & 3529 & 1088 & 1398 & 11595 \\
Test  & 527  & 462  & 407  & 883  & 273  & 350  & 2902  \\
Total & 2633 & 2308 & 2035 & 4412 & 1361 & 1748 & 14497 \\
\hline
\end{tabular}
}
\tablecomments{Details of the dataset used for model training and evaluation. It is based on the Galaxy10 DECaLS dataset after data cleaning and category merging.}
\label{tab:dataset_stat}
\end{table}

The Galaxy10 DECaLS dataset was constructed from the Galaxy Zoo DR2 \citep{2013MNRAS.435.2835W} and the DESI Legacy Imaging Surveys \citep{dey2019overview} to enable large-scale galaxy classification using high-resolution images. Initially, Galaxy Zoo utilized approximately 270,000 galaxy images from the SDSS dataset, with around 22,000 classified into 10 main categories through volunteer contributions. This dataset was subsequently expanded by incorporating high-quality DECaLS images \citep{walmsley2022galaxy}, resulting in an updated dataset of approximately 441,000 unique galaxies. Through volunteer votes and rigorous filtering, $\sim 18,000$ images were categorized into 10 distinct classes: Disturbed Galaxies (1,081 images), Merging Galaxies (1,853), Round Smooth Galaxies (2,645), In-between Round Smooth Galaxies (2,027), Cigar Shaped Smooth Galaxies (334), Barred Spiral Galaxies (2,043), Unbarred Tight Spiral Galaxies (1,829), Unbarred Loose Spiral Galaxies (2,628), Edge-on Galaxies without Bulge (1,423), and Edge-on Galaxies with Bulge (1,873). Although substantially larger datasets are available \citep[e.g.,][]{2013MNRAS.435.2835W}, in this work we deliberately adopt the refined Galaxy10 DECaLS sample. This decision is motivated by three considerations: (i) limited computational resources, which make it infeasible for us to train diffusion models on extremely large datasets; (ii) the feasibility of conducting additional manual data cleaning at this scale, ensuring higher data quality; and (iii) the relatively balanced number of images across categories, which facilitates more reliable training and evaluation.

The categories Disturbed Galaxies and Merging Galaxies were removed from the dataset because their sample sizes were too small to adequately cover the full hypothesis space. These two types of galaxies often exhibit irregular morphologies, leading to substantial diversity within each category, which made it difficult for the model to effectively learn their representations with the available data. We are aware that Disturbed or Merging Galaxies are important. These samples are not included due to sample number limitation and this work is largely a proof-of-concept work.

To optimize the dataset, Cigar Shaped Smooth Galaxies were merged into In-between Round Smooth Galaxies because of their morphological similarity and the small number of samples in the former category. This adjustment ensured more sufficient training data for each class. Additionally, Unbarred Tight Spiral Galaxies and Unbarred Loose Spiral Galaxies were combined into a single category, Unbarred Spiral Galaxies, as their differences in spiral arm tightness were minimal compared to other inter-category features. This merger reduced classification complexity while maintaining sufficient training data for each category.

\begin{figure}[ht!]
    \plotone{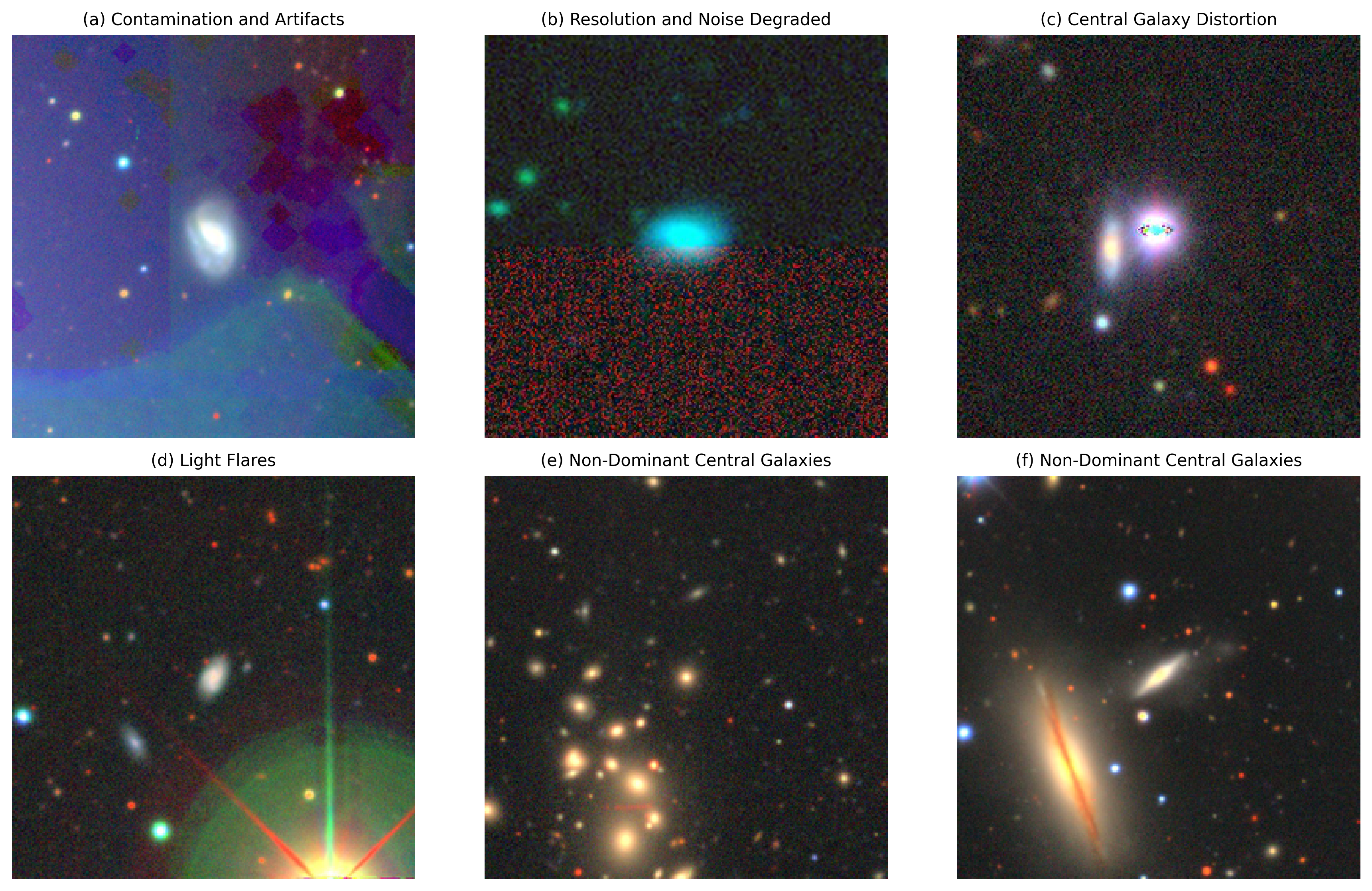}
    \caption{A total of 311 images were excluded from the dataset to improve the quality and category accuracy of the generated results. The images were removed due to factors such as low quality, contamination, or the lack of a dominant central galaxy. For example, (a) images with significant color contamination, (b) images exhibiting excessive noise, (c) the central galaxy’s brightness is distorted, (d) strong light flares in the lower-right corner, (e) multiple galaxies where the central galaxy is not prominent, and (f) the central galaxy is not the main focus and another galaxy in the lower-left corner dominates.
\label{fig:deleteimg}}
\end{figure}

After merging the datasets, further cleaning was performed to ensure data quality. Each image underwent visual inspection, and those with significant distortions—such as excessive contamination, high levels of noise, or prominent bright flares—were removed. Additionally, images where the class labels did not correspond to the central galaxy, which is the largest and most prominent in the figure, were excluded. For instance, images containing multiple galaxies where the central labeled galaxy was not prominent, or those where a larger galaxy adjacent to the labeled center dominated the figure, were eliminated to prevent misleading the model during training. Based on these criteria, a total of 311 images were removed. Examples of the excluded images are shown in Fig.~\ref{fig:deleteimg}.

\begin{figure}[htbp]
    \plotone{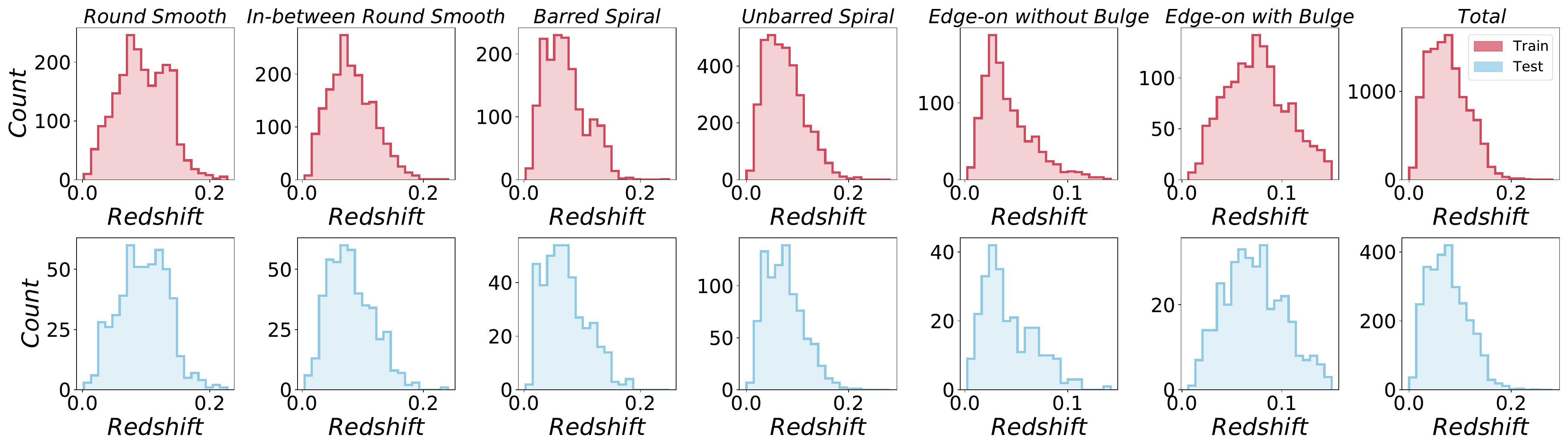}
    \caption{Redshift distribution of the training set and test set for each of the six morphological categories as well as the combined sample. Overall, the distributions are broadly consistent across training and test sets, with minor shifts observed in minority categories. For instance, the peak of the test set distribution for Edge-on with Bulge appears broader compared to that of the training set. Red: training data samples; Blue: testing data samples.
    \label{fig:data_redshift}}
\end{figure}

\begin{figure}[htbp]
    \plotone{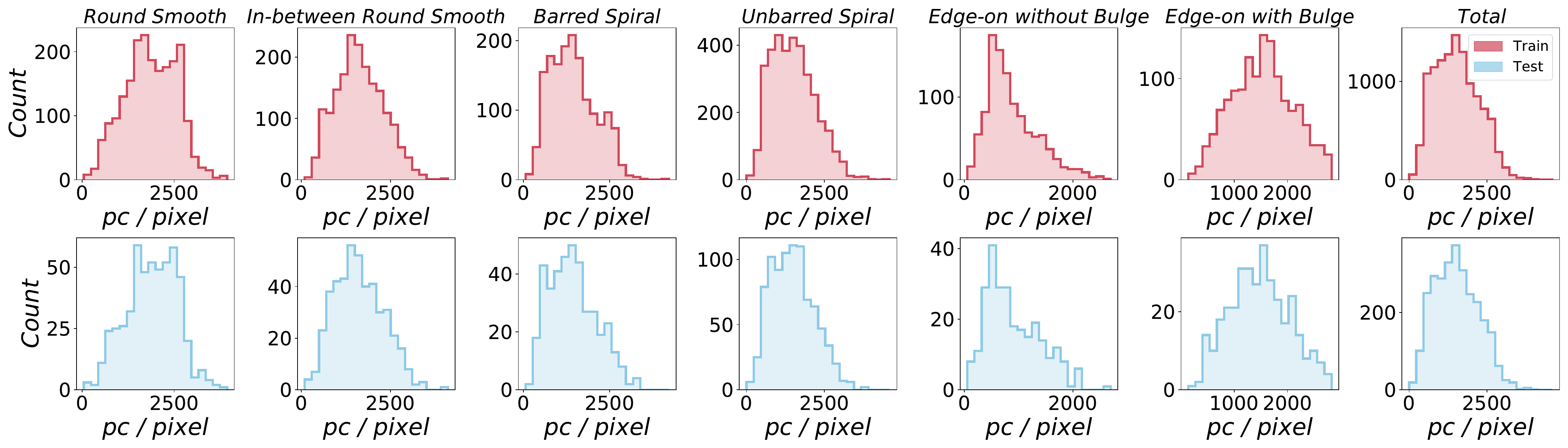}
    \caption{Angular span distribution of the training set and test set for each of the six morphological categories as well as the combined sample. Overall, the distributions are broadly consistent across training and test sets, with minor shifts observed in minority categories. For instance, the tail of the test set distribution for Edge-on without Bulge is relatively higher compared to that of the training set. Red: training data samples; Blue: testing data samples.
    \label{fig:data_scale}}
\end{figure}

After data cleaning, we obtained the final dataset and report the redshift of each galaxy category and its corresponding angular span on the image (pc/pixel), which are presented separately in Fig.\ref{fig:data_redshift} and Fig.\ref{fig:data_scale}. It is worth noting that the redshifts were estimated using a random forest algorithm for photometric redshift prediction, and therefore some values may be subject to uncertainties \citep[e.g.,][]{2021MNRAS.501.3309Z}. Moreover, the training and test sets exhibit broadly consistent distributions in both redshift and angular span, with minor deviations in minority categories (e.g., Edge-on without Bulge). This consistency supports the reliability of the test set as a proxy for assessing the model’s generalization ability.

\subsection{Training Settings}
\label{subsec:trainsetting}

Due to computational resource limitations, the input images in this study were downsampled directly from $256 \times 256 \times 3$ to $64 \times 64 \times 3$, corresponding to the height, width, and number of channels. The image resolution was thus degraded from $0.262 \ \mathrm{arcsec\ pixel}^{-1}$ to $1.048 \ \mathrm{arcsec\ pixel}^{-1}$. Through our early experiments, we manually found that reusing the training parameters from previous studies yields the best performance for our model. We set the batch size to 32 and initialized the learning rate to $10^{-4}$. The Adam optimizer \citep{kingma2014adam} was employed with hyperparameters $\beta_1 = 0.9$ and $\beta_2 = 0.999$, which are the default parameters in PyTorch\footnote{https://pytorch.org/docs/stable/generated/torch.optim.Adam.html}. In our training, we did not apply data augmentation strategies. We tested common augmentations such as random rotations and flips, but observed no noticeable improvements, likely because the dataset already contains galaxies with diverse orientations.

To enhance training stability, the exponential moving average (EMA) was applied to smooth the model parameters \citep{tarvainen2017mean}. The smoothing is computed as
\begin{equation}
    \bar{\theta}_t = \beta \bar{\theta}_{t-1} + (1 - \beta) \theta_t
    \quad ,
\end{equation}
where $\beta = 0.995$ is the smoothing factor, $\theta_t$ represents the current model parameters, $\bar{\theta}_{t-1}$ is the EMA value from the previous step, and $\bar{\theta}_0 = \theta_0$ serves as the initial value. The initial learning rate and the $\beta$ for the EMA were set to the same values as those in \cite{zhao2023can}. The model was implemented using PyTorch \citep{paszke2019pytorch} and trained on a single NVIDIA 3090 GPU.

The loss function (Eq.~\ref{eq:loss}) in diffusion models primarily reflects the model's ability to reconstruct noise.
However, the quality of generated images is determined by a combination of perceptually relevant attributes, such as semantic consistency, texture details, and overall structure. Thus, the loss value alone is insufficient to comprehensively measure the perceptual quality of the generated images \citep{dhariwal2021diffusion}. To address this limitation, we used the Fréchet Inception Distance \citep[FID;][]{heusel2017gans} to evaluate the model's generation quality. FID evaluates the quality and diversity of generated samples by comparing the distribution of generated images to that of real images in the feature space of the Inception-V3 \citep{szegedy2016rethinking}. A  lower FID score indicates higher generation quality. 

During training, FID was calculated for every 5000 model training epochs by generating images equal to the test set. As training progresses, FID typically decreases, reaching its minimum value before slowly and gradually increasing again. By identifying this turning point, we determined the best model. At the same time, category embedding information was randomly dropped with a probability of 10\% as a regularization strategy. This approach encouraged the model to explore the latent features of the data distribution, enhancing the diversity of generated results \citep[e.g.,][]{srivastava2014dropout,ghiasi2018dropblock}.

\section{Results and Discussion} 
\label{sec:result}

\begin{figure}[htbp]
    \plotone{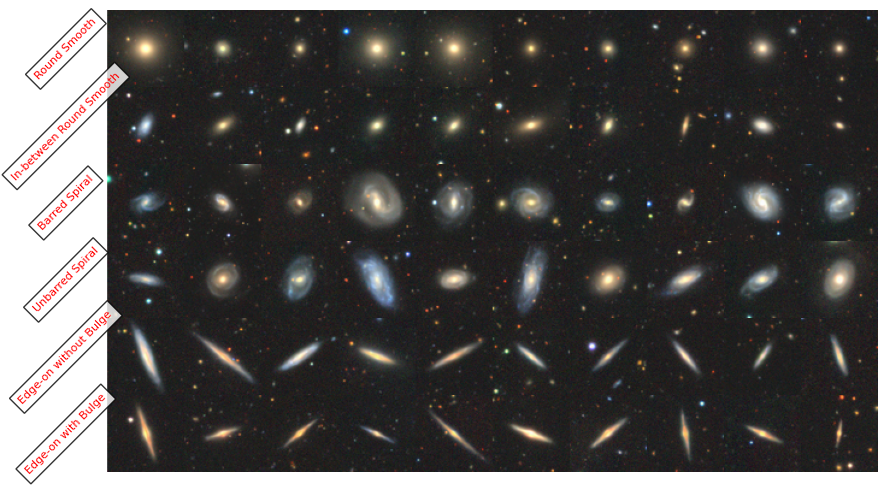}
    \caption{The images were generated by GalCatDiff after 80k training epochs on the Galaxy6 DECaLS dataset. For each of the six galaxy classes, ten images were generated as references. The g, r, and z channels were mapped to the r, g, and b channels. These generated galaxies maintain high image quality while preserving both inter-class and intra-class diversity. Key distinguishing features between galaxy classes are easily identifiable. For instance, the inclination angle is noticeable in Round Smooth and In-between Round Smooth galaxies, while the presence of bars is evident in Barred Spiral and Unbarred Spiral galaxies. The bulge is clearly defined in Edge-on without Bulge and Edge-on with Bulge galaxies. Within each class, galaxies vary in size, shape, and color, yet retain their class-specific characteristics.
    \label{fig:results}}
\end{figure}

This section discusses the quality of images generated by GalCatDiff, with an emphasis on their consistency with galaxy physical properties. After sufficient training, GalCatDiff achieved the optimal FID at the 80k model training epoch. We randomly generated 10 images from each of the six categories. The generated images closely resemble real ones, as shown in Fig.~\ref{fig:results}, and are difficult to distinguish visually.

To provide a comprehensive comparison, we evaluated Astrddpm, an earlier framework that applied diffusion models to galaxy image generation \citep{smith2022realistic}. Since Astrddpm does not support category-specific generation, we embedded category information into Astrddpm using the same method as in our framework to ensure a fair comparison. For GalCatDiff, we conducted an ablation study on the Astro-RAB to evaluate its impact on model performance. Additionally, to mitigate the randomness of a single experiment, we repeated the evaluation five times for each model to better reflect the overall performance.

\subsection{Image Quality}\label{subsec:imagequal}

As discussed in Section~\ref{subsec:trainsetting}, we evaluate image quality using the Fréchet Inception Distance (FID). For the Total FID reported in Table~\ref{tab:FID_scores}, for the unconditional case (Astrddpm), we generated 2,902 images and compared them against the entire test set of 2,902 real images; for the conditional case (all other models), we used class embeddings to generate images, matching the number of samples in each category to that of the test set. These class-wise generated images were then combined, yielding a total of 2,902 samples, which were compared against the full test set. In both cases, the FID was computed once on the entire generated set versus the entire test set, rather than as a per-class average. Thus, the comparison between the unconditional and conditional models is on equal footing. In addition to the Total FID of Astrddpm, Table~\ref{tab:FID_scores} also reports the FID for each category, obtained by using class embeddings to generate the same number of samples as in the test set for that category and then computing FID against its test images. As shown in Table~\ref{tab:FID_scores}, the original Astrddpm shows the least favorable performance. This can primarily be attributed to its inability to capture category variable information. Due to the imbalanced category distribution in the test set and the lack of categorical guidance in Astrddpm, the model struggles to account for quantity discrepancies. As a result, the generated sample distribution deviates from the true test set distribution.
Additionally, the significant feature differences between categories make it difficult for the model to generate the correct category-specific features, further amplifying the performance gap. In contrast, when category embeddings are incorporated into Astrddpm, the FID score improves significantly, dropping from 30.5 to approximately 12. This improvement suggests that the inclusion of category embeddings enables the model to effectively capture category-related information, resulting in generated samples that more closely align with the target distribution.

CatGalDiff shows significant improvements in both the quality of individual categories and overall quality. The improvements for each category are 30.02\%, 13.82\%, 31.81\%, 40.2\%, 20.95\%, and 23.03\%, while the overall FID decreases from 11.7 to 5.3, representing an enhancement of 54.39\%. In the ablation study of the Astro-RAB, we performed a $t$-test on the overall FID, resulting in a p-value of 0.35. This indicates that the improvement in image quality due to the Astro-RAB is not statistically significant. In contrast, improvements to the U-Net architecture significantly enhance image generation quality. We attribute this improvement primarily to the additional skip connections (at least two) introduced at the downsampling and upsampling stages in the U-Net. These connections help the decoder preserve high-resolution information, minimizing detail loss and enabling the model to more effectively capture and reconstruct fine structural details. As a result, noise prediction becomes more precise, leading to a significant enhancement in the overall quality of the generated images. We also observed a positive correlation between the improvement rate and the number of instances per category in the dataset, with a Pearson correlation coefficient of 0.715.

\begin{table}
\centering
\caption{Comparison of FID Scores Across Models and Categories.}
\resizebox{\textwidth}{!}{
\begin{tabular}{lccccccc}
\hline
\textbf{Model} & \textbf{Round Smooth} & \textbf{In-between Round Smooth} & \textbf{Barred Spiral} & \textbf{UnBarred Spiral} & \textbf{Edge-on without Bulge} & \textbf{Edge-on with Bulge} & \textbf{Total FID} \\
\hline
Astrddpm & - & - & - & - & - & - & $30.493 \pm 0.548$ \\
Astrddpm with Class Embedding & $16.290 \pm 0.604$ & $16.271 \pm 0.715$ & $32.208 \pm 0.781$ & $19.946 \pm 0.572$ & $39.485 \pm 0.450$ & $29.853 \pm 0.866$ & $11.735 \pm 0.148$ \\
GalCatDiff without Astro-RAB & $11.936 \pm 0.444$ & $13.815 \pm 0.220$ & $21.908 \pm 0.495$ & $12.903 \pm 0.198$ & $31.464 \pm 0.884$ & $22.598 \pm 0.447$ & $5.420 \pm 0.085$ \\
GalCatDiff & $11.401 \pm 0.181$ & $14.022 \pm 0.313$ & $21.964 \pm 0.338$ & $11.928 \pm 0.089$ & $31.211 \pm 1.005$ & $22.978 \pm 0.561$ & $5.354 \pm 0.120$ \\
\hline
\end{tabular}
}
\tablecomments{Comparisons were made across categories and the full dataset, with the same number of images as the test dataset generated and tested with different models. The Astrddpm model does not support category-specific generation and was only compared on the full test set. Under the GalCatDiff framework, the total FID increased by more than five times compared to Astrddpm. To a fairer comparison, we added class embedding to Astrddpm(baseline), but GalCatDiff still outperformed it by more than two times. Furthermore, Astro-RAB does not result in a statistically significant improvement in image quality.}
\label{tab:FID_scores}
\end{table}


\subsection{Physical properties}
\label{subsec:physprop}

To demonstrate that our model captures measurable galaxy features, we directly compare the galaxy size and flux distributions of generated and real test images. Both sets are evaluated in the same normalized intensity space using identical preprocessing. Since pixel values are commonly normalized to the [$0,1$] range for neural-network training stability \citep[e.g.,][]{isola2017image,zhu2017unpaired, karras2019style, ho2020denoising}, the flux values measured from the generated images should be interpreted as proxies for the true physical fluxes, which we refer to as \textit{pseudo-flux}. Based on the pseudo-flux, we further define a \textit{pseudo-mag} as an analog of aperture photometry in the g-, r-, and z-bands. Specifically, we measure the pseudo-flux by summing the values within a fixed aperture of 3 pixels in diameter (corresponding to $\sim$3 arcsec), and convert it to a pseudo-mag via

\begin{equation}
\mathrm{pseudo\text{-}mag}
= 22.5 - 2.5 \log_{10}\!\left(\mathrm{pseudo\text{-}flux}\right)
\end{equation}

This aperture size is also broadly consistent with the 3-arcsec fibers used in the SDSS spectrographs\footnote{https://www.sdss4.org/dr12/algorithms/magnitudes/} to define quantities such as fiberMag, and therefore provides a physically meaningful and observationally motivated scale for measuring central fluxes. All color indices reported in this work are computed consistently from the pseudo-mag.

We use the Wasserstein-1 distance as an indicator to measure the difference between the generated galaxy size and flux distributions and the test set. The Wasserstein-1 distance is a metric for measuring the difference between two probability distributions, and its core idea is to quantify the minimum amount of work required to transform one distribution into another.
For discrete distributions, suppose $P$ and $Q$ are defined on the sets of points \( \{ x_1, x_2, \dots, x_n \} \) and \( \{ y_1, y_2, \dots, y_m \} \). The Wasserstein-1 distance can be calculated by minimizing the transportation cost. For each pair of points \( (x_i, y_j) \), the transportation cost is given by the absolute distance \( |x_i - y_j| \):
\begin{equation}
    W_1(P, Q) = \min_{\gamma} \sum_{i=1}^{n} \sum_{j=1}^{m} \gamma_{ij} \cdot |x_i - y_j|
    \quad ,
\end{equation}
where \( \gamma_{ij} \) represents the amount of transport flow from point \( x_i \) to point \( y_j \). The goal is to find the optimal transport plan by minimizing the total transportation cost between all possible pairs of points. In practice, we use 
\begin{equation}
    W(u, v) = \int_{-\infty}^{\infty} |U - V|
\label{eq:W1}
\end{equation}
to calculate Wasserstein-1 distance, where U and V are the respective cumulative distribution functions of u and v \citep{smith2022realistic}.

We define a color metric, the \textit{galaxy color Wasserstein distance (Color-WD)}, as the sum of the Wasserstein-1 distances between the pseudo-mag in the g-, r-, and z-bands, and the color index (g-r, r-z) of the generated results and the test set. This can be expressed as:
\begin{equation}
    \begin{array}{r}
        \text{Color-WD} = W_1(G_{\text{g-band}}, T_{\text{g-band}}) + W_1(G_{\text{r-band}}, T_{\text{r-band}}) \\
        + W_1(G_{\text{z-band}}, T_{\text{z-band}}) + W_1(G_{\text{g-r}}, T_{\text{g-r}})+ W_1(G_{\text{r-z}}, T_{\text{r-z}})
    \end{array}
    \quad ,
    \label{eq:Wcolor}
\end{equation}
where $G$ and $T$ represent the generated and test set.


Before defining the size metric (the \textit{half-light radius Wasserstein distance}), we describe how the half-light radius $R_e$ is computed. 
We estimate $R_e$ directly from each normalized galaxy image using a non-parametric procedure consistent with the implementation in the public codebase released by \citet{smith2022realistic}\footnote{\url{https://github.com/Smith42/synthetic-galaxy-distance/}}.
We first measure the median flux within a small central aperture, which provides a stable reference for the central surface brightness. We then expand the radius one pixel at a time and compute the median flux in each newly added annulus, forming a discrete radial brightness profile. The half-light radius is defined as the smallest radius at which the annular median flux drops to 50\% of the central value. Because the radius grows in steps of one pixel, the resulting size estimates are inherently discrete.  After downsampling the images to $64\times 64$ resolution, each pixel corresponds to approximately $1.048$ arcsec (Section~\ref{subsec:trainsetting}). 
Thus, the half-light radii naturally appear quantized at $\sim$1-arcsec intervals.

We then define the size metric, \textit{half-light radius Wasserstein distance} (\(\mathit{HLR\text{-}WD}\)), calculated using Eq.~\ref{eq:W1}. It is given by
\begin{equation}
\mathit{HLR\text{-}WD} = W_{1}(G_{R_e}, T_{R_e}),
\label{eq:WRe}
\end{equation}
where \(W_{1}(G_{R_e}, T_{R_e})\) represents the Wasserstein-1 distance between the generated and test sets of the half-light radius.

We finally define the \textit{weighted synthetic galaxy distance (WSGD)} metric as the average of the \textit{galaxy color Wasserstein distance (Color-WD)} and the \textit{half-light radius Wasserstein distance} (\(\mathit{HLR\text{-}WD}\)):
\begin{equation}
\mathit{WSGD} = \frac{1}{2}\bigl(\mathit{Color\text{-}WD} + \mathit{HLR\text{-}WD}\bigr)
    \quad ,
    \label{eq:WSGD}
\end{equation}

\begin{figure}[htbp]
    \plotone{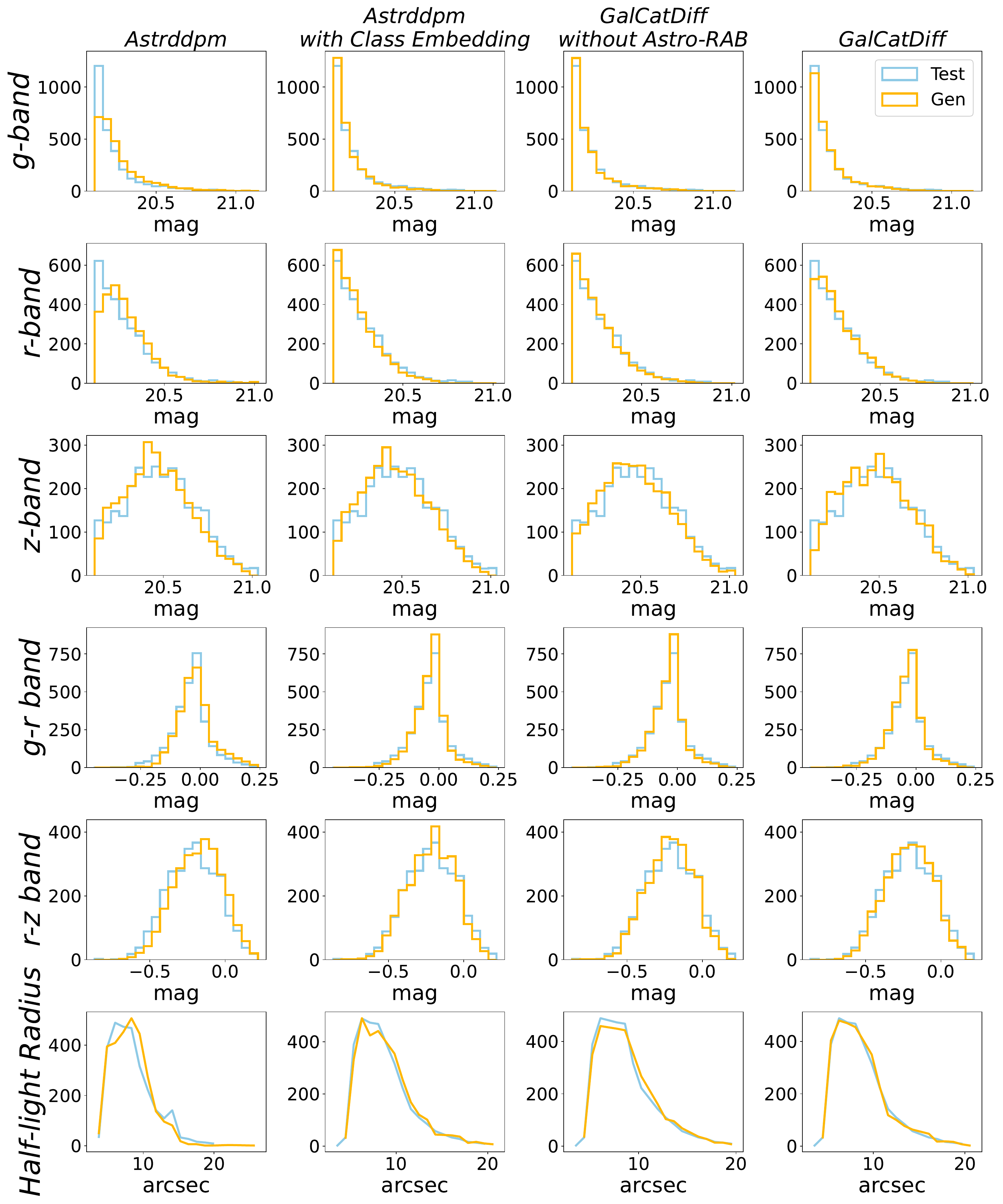}
    \caption{The histograms compare key features—g, r, and z band aperture magnitudes, color index g–r and r–z, and half-light radii—between the images generated by different models (trained on the Galaxy6 DECaLS dataset) and the entire test set (all classes combined). The figure contains 24 distribution plots, organized into 6 rows and 4 columns, with each row representing a different feature and each column corresponding to a specific generative model: Astrddpm, Astddpm with Class Embedding, GalCatDiff without Astro-RAB, and GalCatDiff. Yellow: generated data samples; Blue: testing data samples. For visualization purposes, the half-light radius distributions, discretized at $\sim$1-arcsec intervals, are shown by connecting the histogram bin centers.
\label{fig:phydist}}
\end{figure}

This section evaluates the model’s performance using the entire test dataset. While category-wise evaluation is possible, the total test size (2,902 images) and the uneven distribution across categories (see Table~\ref{tab:dataset_stat}) mean that some classes contain far fewer examples than others. As a result, per-category metrics may be noisy or biased, so we primarily report overall results, with category-level values provided in Table~\ref{tab:model_performance} for reference. As shown in Table~\ref{tab:model_performance}, the Astrddpm model consistently performed the worst among all tested methods. By incorporating class embeddings, we improved the model’s ability to represent the physical properties of galaxies, as discussed in Section~\ref{subsec:imagequal}. Our Enhanced U-Net architecture demonstrated substantial improvements across all metrics, including a 31.65\% increase in weighted synthetic galaxy distance, a 41.83\% improvement in the Wasserstein-1 Distance of the half light radius, and a 39.77\% enhancement in WSGD. Adding the Astro-RAB further amplified these gains, achieving a 25.93\% improvement in the color metric, a 43.14\% increase in the size metric, and a 38.83\% enhancement in WSGD. Importantly, the Astro-RAB also achieved a significant reduction in the standard deviation of results, decreasing the standard deviation of the color metric by 17.67\%, the size metric by 26.32\%, and the WSGD by 45.45\%. In Fig.~\ref{fig:phydist}, we present histograms of color metric (g, r, z, color index) and half-light radius for each model on the entire test set (without class separation). Astrddpm remains the weakest performer: it exhibits substantial discrepancies in peak amplitudes (g-band, z-band), fails to align peak positions with the test set (r-band, r–z band, half-light radius), and shows an overall distributional shift (z-band, r–z band). After incorporating class embeddings, the peak positions in the g-band and r-band are corrected; however, the amplitudes change from severe underestimation to slight overestimation, which may indicate mild overfitting. In contrast, our Enhanced U-Net architecture further narrows the gap between the g-band and r-band distributions and improves the peak shape consistency in the z-band and r–z band. After incorporating Astro-RAB, the generated distributions achieved the best agreement with the test set in terms of color index (g–r, r–z) and half-light radius. Together with the improvements in evaluation metrics, this indicates that Astro-RAB improves the consistency of generated distributions with the test set and strengthens model stability, leading to more robust and reliable generative performance.

\begin{table}
\centering
\caption{Model Performance by Physics Metrics.}
\begin{tabular}{lccc}
\hline
\textbf{Model} & \textbf{\(\mathit{Color\text{-}WD}\)} & \textbf{\(\mathit{HLR\text{-}WD}\)} & \textbf{$WSGD$} \\
\hline
Astrddpm & $0.127 \pm 0.006$ & $0.358 \pm 0.016$ & $0.243 \pm 0.011$ \\
Astrddpm with Class Embedding & $0.079 \pm 0.006$ & $0.263 \pm 0.019$ & $0.171 \pm 0.010$ \\
GalCatDiff without Astro-RAB & $0.054 \pm 0.006$ & $0.153 \pm 0.019$ & $0.103 \pm 0.011$ \\
GalCatDiff & $0.040 \pm 0.005$ & $0.087 \pm 0.014$ & $0.063 \pm 0.006$ \\
\hline
\end{tabular}
\tablecomments{Due to the imbalance and limited number of samples in each class of the test set, the physics metrics—including the galaxy color Wasserstein distance (Eq.~\ref{eq:Wcolor}), half-light radius Wasserstein-1 distance (Eq.~\ref{eq:WRe}), and weighted synthetic galaxy distance (Eq.~\ref{eq:WSGD})—were calculated and compared across different models on the entire test set to assess the model, as shown in Fig.~\ref{fig:phydist}. With contributions from the U-Net architecture enhancement and the addition of Astro-RAB, GalCatDiff surpassed the baseline by a factor three. Class-wise physics metrics of GalCatDiff based on our test set are presented in Table~\ref{tab:galcatdiff_metrics} for reference.}
\label{tab:model_performance}
\end{table}

\begin{figure}[htbp]
    \plotone{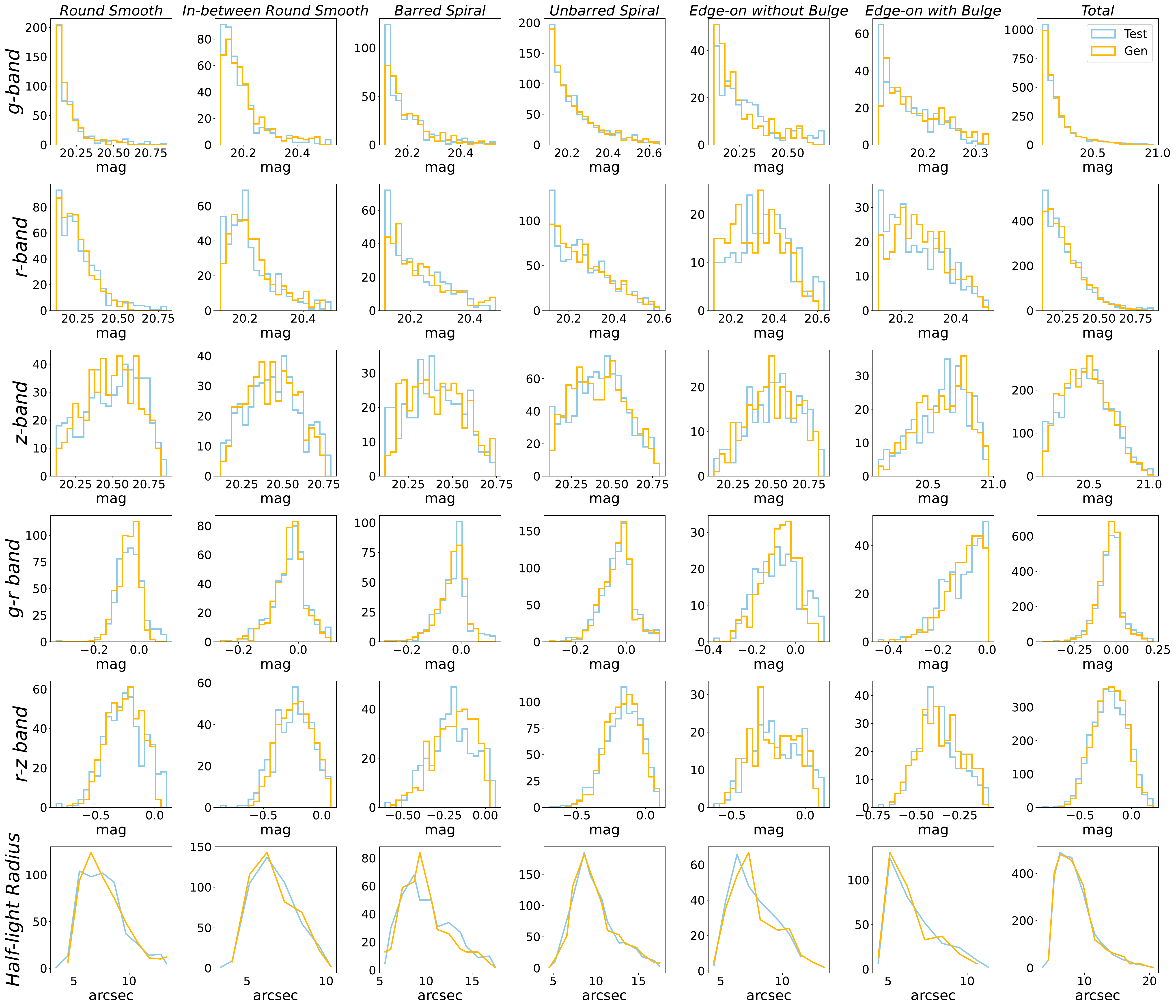}
    \caption{The histograms compare key features—g, r, and z band aperture magnitudes, color index g-r and r-z, and half-light radii—between the images generated by GalCatDiff (trained on the Galaxy6 DECaLS dataset) and the corresponding categories in the test set. The figure contains 42 distribution plots, organized into 6 rows and 7 columns, with each row representing a different feature and each column corresponding to a specific galaxy class, with the rightmost column representing the entire test set (all classes combined). Yellow: generated data samples; Blue: testing data samples. For visualization purposes, the half-light radius distributions, discretized at $\sim$1-arcsec intervals, are shown by connecting the histogram bin centers.
\label{fig:Galcatphydist}}
\end{figure}

\begin{table}
\centering
\caption{GalCatDiff Physics Metrics for Test Set Classes.}
\resizebox{\textwidth}{!}{
\begin{tabular}{lcccccc}
\hline
\textbf{GalCatDiff} & \textbf{Round Smooth} & \textbf{In-between Round Smooth} & \textbf{Barred Spiral} & \textbf{Unbarred Spiral} & \textbf{Edge-on without Bulge} & \textbf{Edge-on with Bulge} \\
\hline
\textbf{\(\mathit{Color\text{-}WD}\)} & $0.095 \pm 0.019$ & $0.065 \pm 0.007$ & $0.082 \pm 0.021$ & $0.036 \pm 0.004$ & $0.115 \pm 0.019$ & $0.071 \pm 0.008$ \\
\textbf{\(\mathit{HLR\text{-}WD}\)} & $0.189 \pm 0.031$ & $0.132 \pm 0.025$ & $0.346 \pm 0.092$ & $0.148 \pm 0.012$ & $0.390 \pm 0.134$ & $0.184 \pm 0.061$ \\
\hline
\end{tabular}
}
\tablecomments{GalCatDiff evaluated the galaxy color Wasserstein distance (Eq.~\ref{eq:Wcolor}) and half-light radius Wasserstein-1 distance (Eq.~\ref{eq:WRe}) on six galaxy classes in the test set. Due to the limited number of samples and the imbalance in class distributions, the generated samples may not fully represent the diverse features of each class. As a result, the evaluation tends to show a higher standard deviation, particularly for minority classes such as Barred Spiral and Edge-on without Bulge.
}
\label{tab:galcatdiff_metrics}
\end{table}

In Fig.~\ref{fig:Galcatphydist}, we present histograms comparing the performance of the GalCatDiff model and the test set across various categories and metrics. The generated distributions generally align with the test set, but some deviations are observed in the tails and peak positions of the distributions. The model's generative performance appears to correlate closely with the number of samples in each category, with categories containing fewer samples often exhibiting poorer generation quality. 

As shown in Table~\ref{tab:galcatdiff_metrics}, the category ``Edge-on without Bulge", which has the fewest samples, performs the worst in terms of the Wasserstein-1 distance for both color and size metrics, and shows the largest standard deviation in size. The histogram for this category reveals an overestimation of the g-band peak and shifts in the peak positions of the r-band and z-band colors. These errors further amplify the deviations in the color indices (g–r and r–z). In addition, a mismatch is observed in the peak of the half-light radius distribution. The category “Barred Spiral,” which has the second-fewest samples in the dataset, performs better than “Edge-on without Bulge.” The g- and r-band peaks are correctly located but differ in height from the test set, and the half-light-radius peak is slightly overestimated. However, it is worth noting that the ``Round Smooth" category ranks second in terms of differences in the color metric. Upon investigation, we found that the dataset contains several extreme galaxies with exceptionally bright centers with different colors, situated at the right-hand tail of the color histogram. These rare samples are challenging for the diffusion model to capture accurately, leading to generated galaxies that are hard to replicate the color features of these outliers.

Despite some discrepancies in specific categories, the overall distribution of the GalCatDiff model aligns well with the test set, particularly showing significant improvements in the matching of peak positions and long-tail distributions for color and size metrics. This indicates that the model effectively captures global features over the full dataset. However, the model performs poorly in a few categories, characterized by shifts in the tails or mismatches in both the peak positions and peak intensities of the distributions. This phenomenon can be attributed to the imbalanced category distribution in the dataset \citep{qin2023class}. Although we deliberately selected a dataset with relatively balanced categories to mitigate this issue, some discrepancies across classes (e.g., between major and minority categories) remain unavoidable and contribute to the observed differences. During training, the model tends to optimize generative performance by learning the data from the dominant categories, thereby capturing the overall feature distribution effectively at a global level. However, because the samples from minority categories contribute less to the overall loss, the model struggles to learn their specific features, resulting in lower generative quality for these categories. In contrast, categories with a larger number of samples contribute more significantly to the loss computation, making the model more attuned to their features. This imbalance leads to inferior generative performance for categories with fewer samples or more complex distributions, as evidenced by deviations in metrics such as color and size.

In addition to category imbalance, intra-category diversity, and extreme samples also significantly impact the model's performance. Most samples in the overall distribution are concentrated in common categories, where the model achieves better feature reconstruction during optimization. However, for extreme samples within minority categories, their rarity limits the model's exposure to them during training, making it challenging to fully capture their characteristics. This intra-category distribution complexity further amplifies the challenge of generating minority categories, resulting in deviations in tail features and peak positions of the generated results. This underscores the difficulty rare samples pose to the model's generative capability.

There are occasional instances where the generated galaxy category does not match the specified category during the generative process. This is primarily due to the impact of data imbalance \citep{qin2023class} and the strategy of discarding category labels with a 10\% probability during training. While this strategy enhances the model's robustness to some extent, it also weakens the supervisory signal for category alignment.


\subsection{Impact of Astro-RAB on Physical Properties}
\label{subsec:impactPPRB}

The Enhanced U-Net architecture and the inclusion of the Astro-RAB both make significant contributions to preserving the physical properties of galaxies in the GalCatDiff framework. We performed two-sample $t$-tests on the WSGD values obtained from multiple independent runs, taking into account their means and standard deviations. One test compared the Enhanced U-Net architecture against the baseline, and another compared the Enhanced U-Net with and without the Astro-RAB. The resulting $p$-values were \(1.94 \times 10^{-8}\) and \(1.66 \times 10^{-4}\), respectively, both far below the common significance level (e.g., \(\alpha = 0.05\)). The performance improvements of the Enhanced U-Net architecture are consistent with the reasons discussed in Section~\ref{subsec:imagequal}, specifically that the additional skip connections help the up-sample stage retain high-resolution information, thereby enhancing the accuracy of noise prediction. Additionally, the improvement in the weighted synthetic galaxy distance and the significant reduction in the standard deviation across all metrics due to the Astro-RAB, as highlighted in Section~\ref{subsec:physprop}, are particularly remarkable and can be attributed to the structural design of the Astro-RAB.

The core of the Astro-RAB lies in its Attention Fusion Unit design, which integrates convolution and window attention mechanisms. Convolutional layers are well-suited for capturing local structures and fine-grained patterns, whereas attention mechanisms excel at modeling long-range dependencies and global context. By leveraging the U-Net to learn weights and output a weighted sum of the convolutional and window-attention features, the module flexibly captures diverse informational features, thereby addressing both local and global characteristics and ensuring physical consistency in the generation of galaxy images.

\begin{figure}[htbp]
    \plotone{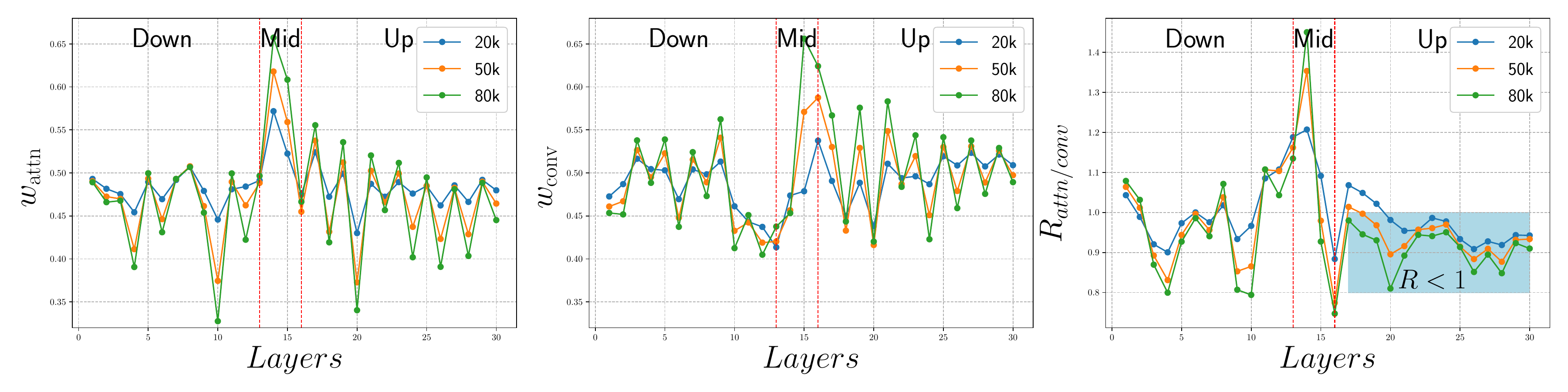}
    \caption{This figure shows the variations of trainable weights across layers during different training stages. The parameters \( w_{\text{attn}} \) and \( w_{\text{conv}} \) are trainable weights from different Attention Fusion Unit layers of GalCatDiff. The output of each Attention Fusion Unit is determined by the weighted sum of the window attention and convolutional layers. The left, middle, and right plots illustrate the variations in \( w_{\text{attn}} \), \( w_{\text{conv}} \), and \( w_{\text{attn}}/w_{\text{conv}} \) across the down-sampling stages, the U-Net bottleneck layer, and the up-sampling stages during the training process at 20k, 50k, and 80k model training epochs. Notably, after sufficient training, \( w_{\text{attn}}/w_{\text{conv}} \) remains consistently below one during the up-sampling stages. During the down-sampling stage and the first two bottleneck layer units, the model captures global relationships to establish global constraints, while the last two bottleneck layer units and the up-sampling stage focus on refining details as local constraints, with both types of constraints working to ensure galaxy physical consistency.
\label{fig:rates}}
\end{figure}

During model training, the weight coefficients for attention and convolution (\( w_{\text{attn}} \) and \( w_{\text{conv}} \)) are gradually adjusted, ultimately reaching an optimal state. Fig.~\ref{fig:rates} illustrates the changes in these weight coefficients for each Attention Fusion Unit from left to right (as shown in Fig.~\ref{fig:unet}) at 20k, 50k, and 80k model training epochs. These changes reveal how the model allocates weights to attention and convolution at different training stages to achieve optimal performance. Experimental results show that the network structure exhibits a distinct stage-dependent pattern: dividing the U-Net at its midpoint, the first half, which includes the down-sampling stage and the first two units of the middle layers, shows significantly higher attention mechanism weights compared to convolution weights in some layers. For instance, in the first Astro-RAB of the middle layer, the attention mechanism weight reaches its peak. At this stage, the model focuses more on capturing the global information of galaxies, such as the overall shape and structure, the gradient of brightness, and the interrelations among different regions. Learning this global information is crucial for diffusion models. In diffusion models, the core objective of the U-Net is to predict the noise initially added by optimizing the loss function (Eq.~\ref{eq:loss}). Therefore, higher attention weights in the earlier stages of the model can better help the network identify the latent structures of galaxies, providing global constraints on galaxy physical consistency.

In the second half of the U-Net, which includes the last two units of the middle layers and the up-sampling stage, convolution weights dominate. As shown in the right panel of Fig.~\ref{fig:rates} starting from layer 15 and throughout the entire up-sampling stage, \( w_{\text{attn}}/w_{\text{conv}} \) consistently remains below one. This indicates that the model focuses more on extracting local features during these stages. The convolution-dominated design of the up-sampling stage demonstrates to be more efficient in restoring the details of galaxy images, particularly excelling in preserving the fine details of core galaxy features. Specifically, during the up-sampling stage, the dominance of convolution enables the model to more accurately capture the local details of galaxies, such as the textures of bulges, bars, or spiral arms. By providing a refined estimation of local structures, the convolution layers offer strong local constraints for noise prediction, allowing the model to more precisely reconstruct the key physical attributes of galaxies. This precise noise estimation helps the model generate galaxy images that are closer to reality while ensuring exceptional performance in terms of physical consistency.

Furthermore, as shown in the right panel of Fig.~\ref{fig:rates}, \( w_{\text{attn}}/w_{\text{conv}} \) in the Attention Fusion Unit of the U-Net bottleneck layer changes significantly. This is because the bottleneck layer serves as a critical transition point in the U-Net's structure, bridging the preceding down-sampling stage with the subsequent up-sampling stage. The dynamic changes in the balance between attention and convolution reflect a gradual shift in feature processing from global to local. During the down-sampling stage, multiple layers of convolution and pooling operations progressively compress the input features to extract high-level information, although some details may be lost in the process. In the early units of the bottleneck layer (layers~13 and~14), the proportion of attention gradually increases, enhancing the model’s ability to capture global information through the attention mechanism. This indicates that the model is fully integrating the global features of galaxies at this stage. In contrast, in the later units of the bottleneck layer (layers~15 and~16), the proportion of attention decreases while the convolution proportion increases, signaling a transition from global information modeling to the recovery of local details. This shift effectively prepares the model for the subsequent up-sampling phase. In summary, these observations suggest that global-to-local integration is accomplished primarily in the bottleneck.

\begin{figure}[htbp]
    \plotone{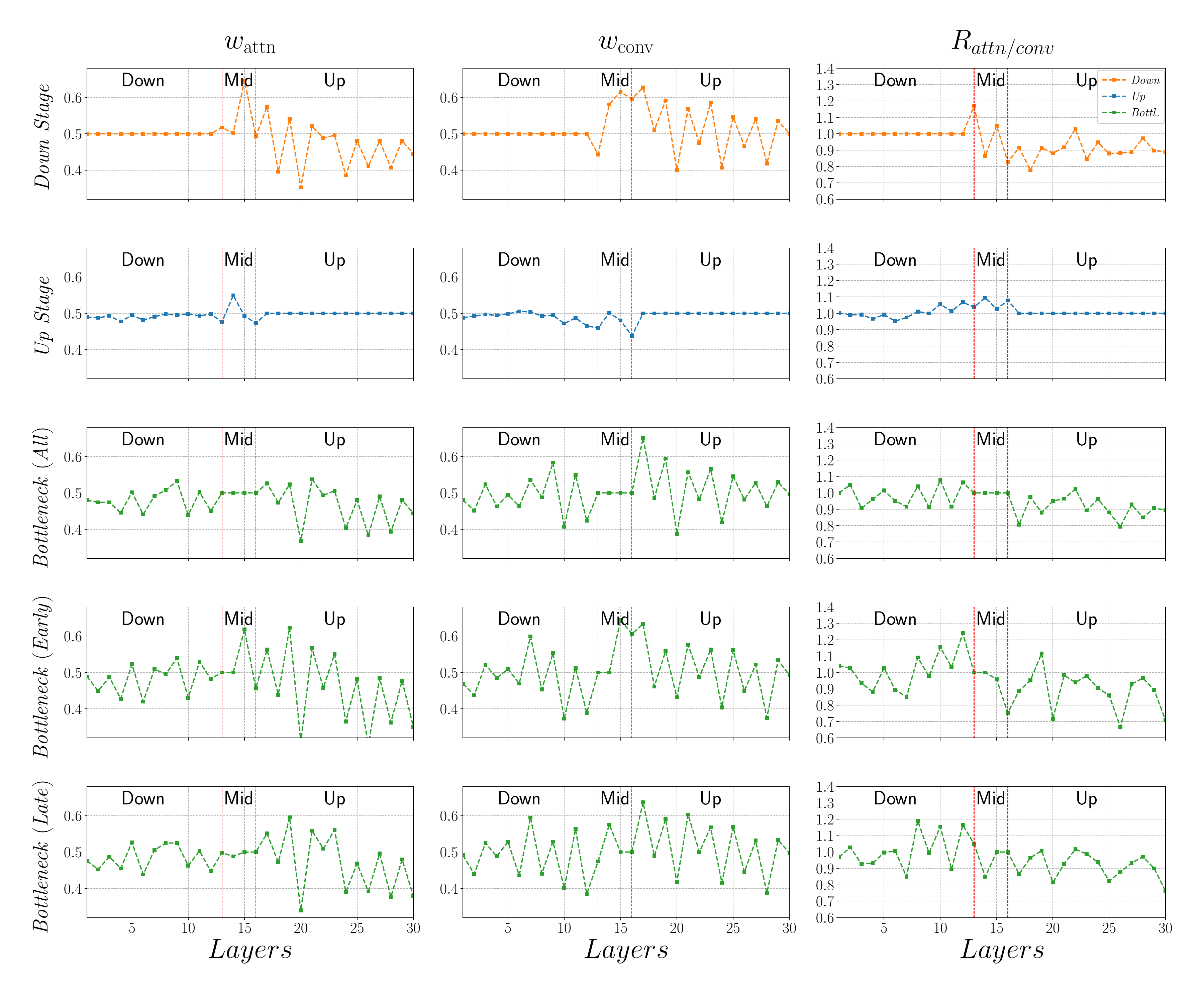}
    \caption{This figure shows the variations of trainable weights across layers under different ablation settings. The parameters \( w_{\text{attn}} \) and \( w_{\text{conv}} \) are trainable weights from different Attention Fusion Unit layers of GalCatDiff. For ablation at a given stage, both \( w_{\text{attn}} \) and \( w_{\text{conv}} \) are fixed to 0.5 and kept constant during training. Each row illustrates layer-level variations variations of these weights across different ablation settings of the U-Net architecture: down-sampling stage, up-sampling stage, and bottleneck stage (all, early, and late ablations), with \( w_{\text{attn}} \), \( w_{\text{conv}} \), and \( w_{\text{attn}}/w_{\text{conv}} \) in different columns. Performance metric variations induced by each ablation are in Table~\ref {tab:stage_performance}.
\label{fig:rates_ablation}}
\end{figure}

To further assess the role of Astro-RAB, we conduct ablation studies on different stages of the U-Net: down-sampling, up-sampling, the bottleneck, the early bottleneck (layers 13 and 14), and the late bottleneck (layers 15 and 16). In each ablation, for the selected stage, we fix the Astro-RAB weights (\( w_{\text{attn}} \) and \( w_{\text{conv}} \)) to 0.5 (\( w_{\text{attn}}/w_{\text{conv}} \) = 1) and keep them constant throughout training. Table.~\ref{tab:stage_performance} reports the evaluation metrics for all ablations, and Fig.~\ref{fig:rates_ablation} shows the per-layer values of (\( w_{\text{attn}} \), \( w_{\text{conv}} \)), and their ratio for the best model under each setting.

As Table.~\ref{tab:stage_performance} indicates, removing Astro-RAB from any stage degrades both the color metrics (Color-WD) and the size metric (HLR-WD). Among all ablations, the up-sampling ablation leads to the minimal degradation, whereas ablating the bottleneck causes pronounced deterioration; in particular, removing Astro-RAB from the late bottleneck (layers 15–16) reduces performance to a level comparable to Astrddpm with class embedding. Taken together, the evidence from the Table.~\ref{tab:stage_performance} and Fig.~\ref{fig:rates} and \ref{fig:rates_ablation} indicates that model performance is closely associated with a global-to-local integration pattern in the bottleneck: attention weights are higher in the early bottleneck (layers 13 and 14), followed by a gradual increase in convolutional weights in the late bottleneck (layers 15 and 16). This configuration first captures global structure and then refines local details, and corresponds to closer agreement with the test-set distributions and lower WSGD. As shown in Fig.~\ref{fig:rates}, the \( w_{\text{attn}}/w_{\text{conv}} \) ratio exhibits a rise-then-fall trajectory across layers 13 to 16; departures from this pattern coincide with marked performance degradation.

Comparative ablations further corroborate this mechanism. In the up-sampling ablation, WSGD increases from 0.063 to 0.075 (+19\%), representing the least degradation; this is plausibly because fixing (\( w_{\text{attn}} \) and \( w_{\text{conv}} \)) at 0.5 provides a reasonable initialization, and the bottleneck still preserves the pattern of higher attention weight in the early bottleneck (e.g., layer 14) followed by increased convolutional weight in the late bottleneck (e.g., layer 15), albeit with smaller amplitude. Ablating the late bottleneck (layers 15 and 16) yields the most severe deterioration, with WSGD increasing from 0.063 to 0.181 (+187\%); in this setting, attention is suppressed in the early bottleneck and only recovers thereafter, while the expected reinforcement of convolutional weights in the late bottleneck is absent, disrupting the intended global-to-local integration. Ablating the entire bottleneck also causes substantial degradation (from 0.063 to 0.141; +124\%). In the down-sampling ablation (from 0.063 to 0.096; +52\%), attention briefly increases in the early bottleneck but quickly declines, failing to establish a stable progression. By comparison, ablating the early bottleneck leads to a milder decline (from 0.063 to 0.089; +41\%) and smaller standard deviations across both metrics, possibly because attention increases during the latter half of down-sampling and convolutional weights strengthen in the late bottleneck, thereby partially preserving the global-to-local integration.

Additionally, in our experiments, the Attention Fusion Unit utilizes an $8\times 8$ pixel window for the window attention mechanism and a $3\times 3$ convolution kernel. This configuration effectively covers the core regions of galaxies, such as the Flux within aperture photometry (1.5-pixel radius), which is likely one of the factors contributing to the significant enhancement of the weighted synthetic galaxy distance.

\begin{table}
\centering
\caption{Model Performance under Ablation Studies}
\begin{tabular}{lcccc}
\hline
\textbf{Model} & \textbf{\(FID\)} & \textbf{\(\mathit{Color\text{-}WD}\)} & \textbf{\(\mathit{HLR\text{-}WD}\)} & \textbf{$WSGD$} \\
\hline
GalCatDiff & $5.354 \pm 0.120$ & $0.040 \pm 0.005$ & $0.087 \pm 0.014$ & $0.063 \pm 0.006$ \\
Down Stage & $5.510 \pm 0.126$ & $0.051 \pm 0.003$ & $0.141 \pm 0.056$ & $0.096 \pm 0.028$ \\
Up Stage & $5.337 \pm 0.060$ & $0.045 \pm 0.001$ & $0.104 \pm 0.037$ & $0.075 \pm 0.019$ \\
Bottleneck (All) & $5.799 \pm 0.179$ & $0.050 \pm 0.006$ & $0.232 \pm 0.036$ & $0.141 \pm 0.020$ \\
Bottleneck (Early) & $5.405 \pm 0.084$ & $0.052 \pm 0.004$ & $0.126 \pm 0.008$ & $0.089 \pm 0.005$ \\
Bottleneck (Late) & $5.581 \pm 0.058$ & $0.072 \pm 0.010$ & $0.291 \pm 0.043$ & $0.181 \pm 0.024$ \\
\hline
\end{tabular}
\tablecomments{The table reports evaluation metrics—including FID, galaxy color Wasserstein-1 distance (Eq.~\ref{eq:Wcolor}), half-light radius Wasserstein-1 distance (Eq.~\ref{eq:WRe}), and weighted synthetic galaxy distance (Eq.~\ref{eq:WSGD})—for different ablation stages of the U-Net architecture. Comparisons highlight the impact of different architectural stages (down, up, and bottleneck) on model performance. Specifically, the bottleneck contains two Astro-RAB modules (each composed of two Attention Fusion Units); Bottleneck (Early) corresponds to ablating the first Astro-RAB, while Bottleneck (Late) corresponds to ablating the second Astro-RAB. Ablating the second Astro-RAB in the bottleneck results in the largest degradation in performance, whereas removing the entire bottleneck yields the next most pronounced decline.}
\label{tab:stage_performance}
\end{table}

\section{Conclusions} 
\label{sec:conclusion}

Galaxy morphology is a frequent focus in astronomical studies as it strongly correlates with physical properties, environmental influences, and the formation and evolutionary history of galaxies. Studies have shown that morphology serves as a powerful tool for investigating galaxy evolution, while generation methods are essential for assessing the consistency between predictions and observations.

Some galaxy generation models have been developed and are capable of generating realistic mock galaxy images, though these methods may require specific assumptions of physics models (semi-analytical models), be unable to generate high-quality images (VAEs), or struggle to perform stable generation (GANs). The most recent method in this field, Astrddpm, made use of advanced diffusion model architecture and can stably generate realistic galaxy mock images \citep{smith2022realistic}. The drawback of such a method, however, is that it requires significant computational resources for model training and has limited capability to benefit category-related downstream tasks, such as fueling the model training of deep learning based galaxy classification models and class-wise galaxy image segmentation models.

In this study, we attempted to overcome these issues in a proof-of-concept manner by developing a diffusion model based galaxy generation framework, GalCatDiff. This framework is ideally able to stably generate realistic galaxy mock images in given classes. We achieved this by firstly developing a novel model training module called Astro-Residual Attention Block (Astro-RAB), a module that combines both attention mechanism and convolution operation to ensure stable local feature reconstruction and galaxy generation. We further designed an enhanced U-Net architecture to optimize model training processes and introduced category embedding to enable class-wise galaxy generation processes.

With the Galaxy6 DECaLS dataset modified from the Galaxy10 DECaLS, we trained a GalCatDiff-based galaxy generation model that can generate realistic mock galaxy images in one of the six given classes (Round Smooth Galaxies, In-between Round Smooth Galaxies, Barred Spiral Galaxies, Unbarred Spiral Galaxies, Edge-on Galaxies without Bulges, and Edge-on Galaxies with Bulges).

We analyze the performance of the GalCatDiff framework in generating high-quality galaxy images, with particular emphasis on its ability to preserve the physical consistency of galaxy properties. After extensive training, GalCatDiff achieved optimal performance after 80,000 iterations. The generated images are visually indistinguishable from real galaxies and effectively capture key physical features.

Through a comparison with the previous Astrddpm framework \citep{smith2022realistic}, we demonstrate the importance of category embeddings. To ensure a fair comparison, we incorporated the same category embedding approach into Astrddpm. The results show that GalCatDiff significantly outperforms Astrddpm in terms of overall image quality and physical consistency. Additionally, an ablation study of the Astro-RAB revealed that, while the module did not result in a statistically significant improvement in image quality, it played a crucial role in maintaining physical consistency and reducing the variance of the generated results.

We also evaluated the consistency between the galaxy images generated by our model and the test set in terms of color and half-light radius distribution using the Wasserstein-1 distance. Compared to Astrddpm with class embedding, GalCatDiff shows improvements of 49.36\% in color metric and 66.92\% in size metric, with reductions in the standard deviation of 16.67\% and 26.32\%. These results further validate the effectiveness of the Astro-RAB. Although some bias remains in the generation of minority classes and extreme samples, the model demonstrates strong overall matching in the distribution of physical metrics.

This paper presents a category-based realistic galaxy image simulation framework GalCatDiff. GalCatDiff can serve as a suboptimal galaxy data augmentor, supplementing data in tasks such as classification, detection, redshift prediction, and morphological analysis. If dynamical equations can be integrated to this framework, GalCatDiff may be able to even generate galaxies with high intra-class diversity (e.g., Disturbed or Merging galaxies) and benefit related studies. Finally, we anticipate GalCatDiff could help addressing gaps in observational data, providing reliable support for astronomical statistics and machine learning applications in the era of Euclid and CSST.


\section*{Acknowledgments}

We thank the anonymous reviewers for their constructive comments and suggestions. This work was supported by the National Natural Science Foundation of China (grant No. 12503111). This work was also supported by the XJTLU Research Development Fund (grant No. RDF-24-02-057). H.M.T. gratefully acknowledges support from the Department of Physics of Xi’an Jiaotong-Liverpool University, the longlasting support from the DoA Tsinghua TAGLAB and machine learning group, and the JBCA machine learning group. X.Z.F. gratefully acknowledges helpful discussions with Prof. Dennis Zaritsky, as well as assistance from Ms. Yuning Zhu in preparing the model architecture diagram.


\section*{Data Statement}

The dataset used in this work can be accessed on request and was derived from the Galaxy Zoo DECaLS dataset \citep{10.5281/zenodo.4196266}. The core code underlying this work is available in Zenodo \citep{10.5281/zenodo.18897649}. A developmental copy is availabe in GitHub\footnote{\url{https://github.com/stevefxz/GalCatDiff/tree/main}}. We hope this code can be useful for future developers training their generative models.

pdflatex sample631.tex
bibtext sample631
pdflatex sample631.tex
pdflatex sample631.tex

\clearpage
\bibliographystyle{aasjournal}
\bibliography{ref}


\end{document}